\documentclass[a4paper,12pt,dvips]{article}
\usepackage{amsmath,amssymb,exscale}   
\usepackage{array,multicol}
\usepackage{afterpage,float,flafter}   
\usepackage{epsfig,rotating,pifont,fancybox}   

\makeatletter   
\@addtoreset{equation}{section}   
\makeatother   
   
\newcommand{\bea}{\begin{eqnarray}}   
\newcommand{\eea}{\end{eqnarray}}   
 
\newcommand{\hone}{\varepsilon^{H_1}}
\newcommand{\htwo}{\varepsilon^{H_2}}
\newcommand{\hi}{\varepsilon^{H_i}}
\newcommand{\lL}{\varepsilon^{\ell_L}}
\newcommand{\eR}{\varepsilon^{e_R}}
\newcommand{\qL}{\varepsilon^{q_L}}
\newcommand{\uR}{\varepsilon^{u_R}}
\newcommand{\dR}{\varepsilon^{d_R}}  
\newcommand{\lLi}{\varepsilon^{\ell_{i\, L}}}   
\newcommand{\eRi}{\varepsilon^{e_{i\, R}}}
\newcommand{\qLi}{\varepsilon^{q_{i\, L}}}
\newcommand{\uRi}{\varepsilon^{u_{i\, R}}}
\newcommand{\dRi}{\varepsilon^{d_{i\, R}}}  
\newcommand{\lLone}{\varepsilon^{\ell_{1\, L}}}

\newcommand{\lLtwo}{\varepsilon^{\ell_{2\, L}}}

\newcommand{\NPB}[3]{\emph{ Nucl.~Phys.} \textbf{B#1} (19#2) #3}   
\newcommand{\PLB}[3]{\emph{ Phys.~Lett.} \textbf{B#1} (19#2) #3}   
\newcommand{\PRD}[3]{\emph{ Phys.~Rev.} \textbf{D#1} (19#2) #3}   
\newcommand{\PRL}[3]{\emph{ Phys.~Rev.~Lett.} \textbf{#1} (19#2) #3}

\newcommand{\RMP}[3]{\emph{ Rev.~Mod.~Phys.} \textbf{#1} (19#2) #3}

\newcommand{\EPJC}[3]{\emph{ Eur.~Phys.~J.} \textbf{C#1} (19#2) #3}

\title{   
\vspace*{-0.8cm}   
\begin{flushright}   
\normalsize{      
IEM-FT-198/99\\
IFT-UAM/CSIC-99-42\\ 
CERN-TH/99-332}
\end{flushright}    
\vspace{.8cm}
\Large\textbf{Electroweak and Flavor Physics in Extensions of the 
Standard Model with Large Extra Dimensions~\footnote{Work 
supported in part by CICYT, Spain, under contract AEN98-0816 and AEN98-1093.}}
\vspace*{.5cm}
\author{\large\textbf
{A.~Delgado$\,^a$, A.~Pomarol$\,^{b,}\,$\footnote{On leave of absence from    
IFAE, Universitat Aut{\`o}noma de Barcelona, E-08193 Bellaterra, Barcelona.}\,
  and M.~Quir{\'o}s}$\,^a$\\ \\
$^a$\emph{Instituto de Estructura de la Materia (CSIC)}\\
\emph{Serrano 123, E-28006 Madrid, Spain}\\ \\
$^b$\emph{Theory Division, CERN}\\
\emph{CH-1211 Geneva 23, Switzerland}}}
\date{}   
\begin{document}
\maketitle

\begin{abstract}
\vspace*{.5cm}
We study the implications of extra dimensions of size
$R\sim 1/$TeV 
on electroweak and flavor physics due to the presence
of Kaluza-Klein excitations of the SM gauge-bosons.
We consider several scenarios with
the SM fermions either living
 in the bulk or being localized at different points
of an extra dimension.
Global fits to electroweak 
observables provide
lower bounds 
on $1/R$, which are generically in the $2$--$5$ TeV range.
We find, however, certain models where the 
fit to electroweak observables is 
better than in the SM, because of an improvement in the prediction to the 
weak charge $Q_W$.
We also consider  the case of softly-broken supersymmetric theories
and we find new non-decoupling effects that put new constraints on $1/R$.
If
quarks of different families live in different points of the extra
dimension,
we find that the
Kaluza-Klein modes of the SM gluons generate (at tree level)
dangerous  flavor and CP-violating interactions.
The lower bounds on $1/R$ can increase in this case up to 
$5000$ TeV, disfavoring  these scenarios in the context of TeV-strings.

\end{abstract}
\vspace{1.cm}   
\begin{flushleft}   
CERN-TH/99-332\\
November 1999
\end{flushleft}   
\newpage

\section{Introduction} 
\label{introduction}

The existence of extra dimensions seems to be a crucial ingredient
in the unification of gravity with the gauge forces.
This is the case, for example, of string theory, where more than 
three spatial dimensions seem to be necessary for the consistency of the 
theory. It has often been assumed that these extra dimensions are compact, 
with radii of Planckean size, $R\sim 10^{-33}$ cm,
and therefore irrelevant for phenomenological purposes.

Nevertheless, it has  recently been suggested that these extra dimensions
could, in fact, be very large and testable in present-day experiments.
The first possibility of lowering the compactification scale to 
$\sim$TeV was given in Ref.~\cite{ignatios}. A different, and 
more drastic, possibility, suggested in Refs.~\cite{lykken,add},
is to lower the fundamental (string) scale down to the electroweak scale.
This scenario requires either very large extra dimensions,
in which only gravity propagates~\cite{add}, 
or a very small string coupling~\cite{pioline}.
Having the fundamental (string) scale close to the TeV implies that
the extra dimensions must appear at energies $\sim$ TeV
(or below for the case of gravity),
and therefore such a possibility can be tested in future colliders.

In this paper we want to study the implications of 
an extra TeV$^{-1}$ dimension. We will consider the case of the 
Standard Model (SM) gauge bosons propagating in this extra dimension,
and analyze the implications of their Kaluza-Klein (KK) excitations
for electroweak and flavor processes.
The effects of these KK excitations are of two kinds.
First, the KK excitations can mix with the $Z$ and $W$ boson of the 
SM and, consequently, modify their masses and couplings.
Second, the KK excitations can
induce new four-fermion operators in the SM lagrangian.
All these effects are in principle large, since
they arise at the tree level. 
Using the fact that the experimental data seem to be 
in excellent agreement with the SM, we will be able to put
bounds on the compactification scale $M_c$. 
From the high-precision electroweak data,
we obtain $M_c\gtrsim 2$--$5$ TeV (depending on the model).
Since only the weak charge, $Q_W$, 
measured in atomic parity violation experiments, seems 
to depart by a few sigmas from the SM prediction~\cite{bennett,qw},  
we will briefly discuss how an extra dimension can improve the SM fit.
KK excitations can also induce flavor-violating interactions if 
fermions of different families 
live at different points of the extra dimension, or if some of them are
localized at fixed points and others live in the bulk of the extra 
dimension. We find that this possibility suffers from severe bounds on
$M_c$, which are in certain cases as large as $M_c\gtrsim 5000$ TeV.

Part of the analysis done here has previously been considered in the 
literature \cite{ny,mp,rw,scc}. Our analysis, however, aims to be 
the most general one and thus it incorporates new scenarios not 
considered before.
For example, we will allow some of the SM fermions to live in the bulk
of the extra dimension,  
or to be localized at different points of the extra dimension.
As we will see, these possibilities offer a new and rich
phenomenology, especially concerning flavor physics.
We will also consider the case in which the higher dimensional theory 
is supersymmetric and yields, after compactification to 4D, the 
minimal supersymmetric extension of the SM (MSSM). 
This case had not been analyzed before and
we will show that, surprisingly, there are tree-level effects arising 
from integrating out the superpartners. 
The reason of these effects is the existence of
a scalar SU(2)$_L$-triplet (required by supersymmetry in five dimensions)
that gets a vacuum expectation value (VEV) 
of the order of $m^2_W/M_c$, 
and modifies the relation between the $Z$ and $W$ masses.

The outline of this paper is as follows.
In section~\ref{effective} we consider the SM in 5D and compute
the effective 4D theory obtained after integrating out the KK excitations 
of the SM gauge fields, assuming the most general framework where the SM
fermions can live either at fixed points or in the bulk of the extra dimension.
In section~\ref{susy} we perform a similar exercise for the case where the 5D
theory is supersymmetric. We pay particular attention to the modification
of the effective theory generated  by the VEV of
the 5D gauge superpartner.
This VEV is always non-zero whenever
a Higgs field acquires a VEV at the 
boundary of the extra dimension.
In section~\ref{electroweak} we compute the
predictions of the theory for different electroweak observables, and
present the corresponding bounds on $M_c$ in section~\ref{bounds}.
In section~\ref{flavor} we consider the 
effect of the KK excitation on flavor observables 
such as $\Delta m_K$, $\varepsilon_K$ and 
$\varepsilon^\prime_K$, and derive new bounds on $M_c$. 
Section~\ref{conclusion} is devoted to our conclusions.

A word of caution  on our calculation is in order. 
Since we are working with gauge theories in more than four
dimensions, one could be worried about the validity of the
perturbative expansion of the theory. Higher dimensional theories are known to 
be strongly coupled at energies above the compactification scale $M_c$,
implying a cutoff at $\sim 10$--$100$ $M_c$. 
This means that we can only trust the 
effect of the first $10$--$100$ KK excitations. 
The effect of the heavy $n$ KK modes ($n>10$--$100$), 
although not calculable,
is estimated to be $\lesssim 10\%$  of  the light KK-modes effect.
The  uncertainty of the effect of an extra dimension
in low-energy processes, $p\sim M_Z< M_c$, 
is thus lower than $10^{-1}\, M^2_Z/M_c^2\sim 10^{-3}\, ({\rm TeV}/M_c)^2$. 
Similarly stringy effects will be small if the string scale is at least  
one order of magnitude larger than the compactification radius as we will 
consider here.

\section{The five-dimensional SM}
\label{effective}

The model we want to study is based on a simple extension of the 
SM to 5D \cite{alex}. The  fifth dimension $x_5$ is compactified on the 
orbifold $S^1/\mathbb{Z}_2$, a circle of radius $R$ 
with the identification $x_5\rightarrow -x_5$.
This is a segment of length $\pi R$ with two 4D boundaries, one at
$x_5=0$ and another at $x_5=\pi R$ 
(the  two fixed points of the orbifold).
This type of compactification is needed  to get chiral fermions.
The SM gauge  fields live in the 5D bulk,
while  the SM fermions, $\psi$, and the Higgs doublets, 
$H_i\ (i=1,2)$~\footnote{We will consider here the SM with two Higgs doublets,
as in the MSSM, to encompass the two different possibilities where the Higgs
VEV is either in the 5D bulk or on the 4D boundary. The case
of a single Higgs is easily recovered when only one of the Higgses 
acquires a VEV. }, can either live in the bulk
or be localized on the 4D boundaries. In this section we will assume that all 
the localized fields live on the $x_5=0$ boundary. We leave to 
section~\ref{flavor} the possibility to have localized fields living in 
different points of the extra dimension.

The  5D lagrangian is given by
\begin{eqnarray}
{\cal L}_{5D}&=&-\frac{1}{4g^2_5}
F_{MN}^2+
\sum_{i}
\Big[(1-\varepsilon^{H_i})
|D_M H_{i}|^2+(1-\varepsilon^{\psi_i})
i\bar \psi_{i}\Gamma^M D_M \psi_{i}\Big]\nonumber\\
&+&\sum_{i}
\Big[\varepsilon^{H_i}|D_\mu H_{i}|^2+
\varepsilon^{\psi_i}i\bar \psi_{i}\sigma^\mu D_\mu \psi_{i}
\Big]
\delta(x_5)\, ,
\end{eqnarray}
where we have introduced the operator $\varepsilon$ defined as
$\varepsilon^{F}=1\ (0)$ for the $F$-field living in the boundary (bulk);
$D_M=\partial_M+iV_M$, $M=(\mu,5)$, and $g_5$ is the 5D gauge coupling.
The fields living in the bulk can be defined to be even or odd under the 
$\mathbb{Z}_2$-parity, i.e. $\Phi_{\pm}(x_5)=\pm\Phi_{\pm}(-x_5)$. 
They can be Fourier-expanded as
\begin{eqnarray}
\Phi_+(x_\mu,x_5)&=&\sum^{\infty}_{n=0}
\cos\frac{nx_5}{R}\Phi^{(n)}_+(x_\mu)\, ,\nonumber\\
\Phi_-(x_\mu,x_5)&=&\sum^{\infty}_{n=1}
\sin\frac{nx_5}{R}\Phi^{(n)}_-(x_\mu)\, ,
\label{fourier}
\end{eqnarray}
where $\Phi^{(n)}_{\pm}$ are the KK excitations of the 5D fields.
Gauge and Higgs bosons living in the 5D bulk 
will be assumed to be even under the $\mathbb{Z}_2$.
Their (massless) zero modes will correspond to the 
gauge and Higgs  SM fields.
Fermions in 5D have two chiralities, $\psi_L$ and $\psi_R$, that can
transform as even or odd under the $\mathbb{Z}_2$. The precise assignment
is a matter of definition. We will choose the even assignment for the
$\psi_L$ ($\psi_R$) components of 
 fermions $\psi$, which are doublets (singlets)
under SU(2)$_L$. As a consequence only the $\psi_L$ 
of SU(2)$_L$ doublets and $\psi_R$ of SU(2)$_L$ singlets have zero modes  
(see Eq.~(\ref{fourier})) and the massless fermion sector is chiral.

Using Eq.~(\ref{fourier}) and integrating  over the fifth dimension, we 
can easily obtain the theory in 4D. This will contain the SM fields
plus their  KK excitations. In order to study the impact of this theory in the 
electroweak and flavor processes, we will integrate out the 
KK excitations at the tree level and at the first order in the 
expansion parameter $X$ defined as~\footnote{As we said in 
the introduction (see also section~\ref{flavor}), 
only the first $10$--$100$ KK
excitations should be considered in the sum $\sum_n 1/n^2$.
Nevertheless, since these modes 
already give 
more than 90$\%$ of the sum, 
we will be considering the full KK tower.}
\begin{equation}
\label{equis}
X=\sum_{n=1}^\infty \frac{2}{n^2}\, \frac{m_Z^2}{M_c^2}=\frac{\pi^2}{3}
\frac{m_Z^2}{M_c^2}\, ,
\end{equation}
where $M_c\equiv 1/R$. As we will see, this approximation is good enough 
since $X$ will be constrained to be very small.

To obtain the effective 4D theory, we only need to take into consideration 
the couplings of the SM fermions to the KK excitations of the electroweak 
gauge bosons and the mass terms of these latter. Notice that Higgses living 
on the boundary will induce mixing terms between the SM gauge bosons and
their KK excitations (this is due to the breaking of $x_5$-translational 
invariance of the boundaries). We therefore define for later use the 
effective mixing angle~\footnote{We are using the notation 
$c_\alpha\equiv\cos\alpha$, $s_\alpha\equiv\sin\alpha$, and so on.}
\begin{equation}
s^2_\alpha=\varepsilon^{H_2}\,s^2_\beta+\varepsilon^{H_1}\,c^2_\beta
\, ,
\label{alpha}
\end{equation}
where, as usual, we have introduced the mixing angle $\beta$
defined as $\tan\beta=\langle H_2\rangle/\langle H_1\rangle$,
with $v^2\equiv\langle H_1\rangle^2+ \langle H_2\rangle^2$ and
$v\simeq 174$ GeV.

\subsection*{The charged electroweak sector}

In the charged sector the 4D lagrangian can be written 
as~\footnote{We work in the unitary gauge~\cite{alex}.}
\begin{equation}
\label{charged}
\mathcal{L}^{ch}=\sum_{a=1}^2 \mathcal{L}_a^{ch}
\end{equation}
with
\begin{eqnarray}
\label{chargeda}
\mathcal{L}_{a}^{ch}&=&
\frac{1}{2}m^{2}_W\left\{
W_a\cdot W_a
+2\sqrt{2}s^2_\alpha\sum_{n=1}^\infty
W_a\cdot W_a^{(n)}\right\}
+\frac{1}{2}M_c^2\sum_{n=1}^\infty \, n^2 \, W_a^{(n)}\cdot W_a^{(n)}
\nonumber\\
&-&g\,W_a\cdot J_a-g\,\sqrt{2} J^{KK}_a\cdot \sum_{n=1}^\infty W_a^{(n)}\, ,
\end{eqnarray}
where $m^{2}_W=g^2v^2/2$, the weak angle $\theta$ is defined by 
$e=g\, s_\theta=g'\, c_\theta$, while the currents are
\begin{eqnarray}
\label{currents}
J_{a\mu}&=&\sum_\psi \bar{\psi}_L \gamma_\mu \frac{\sigma_a}{2}\psi_L\, ,
\nonumber\\
J_{a\mu}^{KK}&=& \sum_\psi \varepsilon^{\psi_L}\bar{\psi}_L \gamma_\mu 
\frac{\sigma_a}{2}\psi_L\, .
\end{eqnarray}
For momenta $p^2\sim m_W^2\ll M_c^2$ we can integrate out 
the KK modes $W_a^{(n)}$ using
their equations of motion and neglecting their kinetic
terms. They yield
\begin{equation}
\label{KKW}
W_a^{(n)}=\frac{\sqrt{2}}{n^2\,M_c^2}\left[-
s^2_\alpha 
m_W^{2}\, W_a+g\, J_a^{KK}\right]+\mathcal{O}(X^2)\, .
\end{equation}
Replacing the solution (\ref{KKW}) into (\ref{chargeda}), we obtain
\begin{equation}
\label{effcharged}
\mathcal{L}^{ch}_{a,\,eff}=\frac{1}{2} M^{2}_W
W_a\cdot W_a -g W_a\cdot\left[J_a-s^2_\alpha\, c_\theta^2
\,X\,  J_a^{KK}\right]
-\frac{g^2}{2\, m_Z^2}\,X\,J_a^{KK}\cdot J_a^{KK}\, ,
\end{equation}
where 
\begin{equation}
\label{mWX}
M_W^2=\left[1-s^4_\alpha c^2_\theta X\right]\, m_W^2\, .
\end{equation}
Finally, for very low momenta, much smaller than the weak scale 
$p^2\ll M_W^2$, the $W_a$ gauge bosons can also be integrated out from the
lagrangian (\ref{effcharged}). The result can be written as
\begin{equation}
\label{lowcharged}
\mathcal{L}^{ch}_{a,\, low}=-\frac{g^2}{2 M_W^2}\left\{
\left[J_a-s^2_\alpha\,c_\theta^2\,
X\,  J_a^{KK}
\right]^2
+\, c_\theta^2\,X\,J_a^{KK}\cdot J_a^{KK}\right\}\, .
\end{equation}
We can use the lagrangian (\ref{lowcharged}) to describe the $\mu$ decay,
from where the Fermi constant is found to be
\begin{equation}
\label{fermi}
\frac{G_F}{\sqrt{2}}=\frac{g^2}{8 M_W^2}
\Big[1+\{\lLone\lLtwo
-(\lLone+\lLtwo)s^2_\alpha \}c^2_\theta X \Big]\, .
\end{equation}
If we assume that the KK modes do not spoil lepton universality 
(as we will do in this section), 
Eq.~(\ref{fermi}) can be written as
\begin{equation}
\label{fermiun}
\frac{G_F}{\sqrt{2}}=\frac{g^2}{8 M_W^2}
\left[1+\lL c_{2\alpha} c^2_\theta 
 X \right]\, .
\end{equation}

\subsection*{The neutral sector}

In the neutral sector the 4D lagrangian can be similarly written as
\begin{eqnarray}
\label{neutral}
\mathcal{L}^{neutral}&=&\frac{1}{2}m_Z^2\,\left\{ Z\cdot Z 
+2\sqrt{2}s^2_\alpha 
\sum_{n=1}^\infty Z\cdot Z^{(n)}\right\}
+\frac{1}{2}M_c^2\sum_{n=1}^{\infty}n^2\,\left[Z^{(n)}\cdot Z^{(n)}+
A^{(n)}\cdot A^{(n)}\right]\nonumber\\
&-&\frac{e}{s_\theta c_\theta}\left[Z\cdot J_Z+\sqrt{2}\sum_{n=1}^\infty
Z^{(n)}\cdot J_Z^{KK}\right]
-e\left[A\cdot J_{em}+\sqrt{2}\sum_{n=1}^\infty A^{(n)}\cdot J_{em}^{KK}
\right]\, ,
\end{eqnarray}
where $m^{2}_Z=(g^2+g^{\prime\, 2})v^2/2$ and the currents are
\begin{eqnarray}
\label{currentsn}
J_{\mu\, Z}&=&\sum_\psi \bar{\psi}\,\gamma_\mu 
\left(g_{V}^{\psi}+\gamma_5\,
g_{A}^{\psi} \right)\psi\, ,\nonumber\\
J_{\mu\, Z}^{KK}&=&\sum_\psi \bar{\psi}\,\gamma_\mu 
\left(g_{V}^{\psi, KK}+\gamma_5\,
g_{A}^{\psi, KK} \right)\psi\, ,\nonumber\\
J_{\mu\, em}^{KK}&=&\sum_\psi \bar{\psi}\,\gamma_\mu 
\left(g_{em,V}^{\psi, KK}+\gamma_5\,
g_{em,A}^{\psi, KK} \right)\psi\, .
\end{eqnarray}
The vector and axial couplings are defined by
\begin{eqnarray}
\label{kkcouplings}
g_V^{\psi}&=& \frac{T_3}{2}-s^2_\theta\ Q\, ,\nonumber\\
g_A^{\psi}&=&-\frac{T_3}{2}\, ,\nonumber\\
g_V^{\psi,\,KK}&=&\varepsilon^{\psi_L}\left(\frac{T_3}{2}
-s^2_\theta\,\frac{Q}{2}
\right)-\varepsilon^{\psi_R}s^2_\theta\,\frac{Q}{2}\, ,\nonumber\\
g_A^{\psi,\,KK}&=&\varepsilon^{\psi_L}\left(-\frac{T_3}{2}+s^2_\theta\,
\frac{Q}{2}\right)-\varepsilon^{\psi_R}s^2_\theta\,\frac{Q}{2}\, ,\nonumber\\
g_{em,\,V}^{\psi,\,KK}&=&\left(\varepsilon^{\psi_R}+\varepsilon^{\psi_L}
\right)\, \frac{Q}{2}\, ,\nonumber\\
g_{em,\,A}^{\psi,\,KK}&=&\left(\varepsilon^{\psi_R}-\varepsilon^{\psi_L}
\right)\, \frac{Q}{2}\, .
\end{eqnarray}
For momenta $p^2\ll M^2_c$ the $Z^{(n)}$ and $A^{(n)}$ modes can be 
integrated out, yielding the effective lagrangian
\begin{eqnarray}
\label{effneutral}
\mathcal{L}^{neutral}_{eff}&=&\frac{1}{2} M_Z^2\, Z\cdot Z 
-\frac{e}{s_\theta\, c_\theta}\, Z\cdot\left[J_Z-\,
s^2_\alpha X\,  J_Z^{KK}
\right]-e A\cdot J_{em}
\nonumber\\
&-&\frac{1}{2\,M_Z^2}\,\frac{e^2}{s^2_\theta\, c^2_\theta}\,X
J_Z^{KK}\cdot J_Z^{KK}
-\frac{e^2}{2\, M_Z^2}\,X\, J_{em}^{KK}\cdot J_{em}^{KK}\, ,
\end{eqnarray}
where we have defined
\begin{equation}
\label{mZX}
M_Z^2=
\left[1-s^4_\alpha X \right]m_Z^2\, .
\end{equation}
For momenta $p^2\ll M_Z^2$ we can integrate out the SM 
$Z$-boson and obtain the
neutral lagrangian 
\begin{eqnarray}
\label{lowneutral}
\mathcal{L}^{neutral}_{low}&=&-\,\frac{1}{2 M_Z^2}\,
\frac{e^2}{s^2_\theta\, c^2_\theta}\,
\left\{
\left[J_Z-
s^2_\alpha
X\,  J_Z^{KK}\right]^2
-\,X\,J_Z^{KK}\cdot J_Z^{KK}\right\}\nonumber\\
&-&e A\cdot J^{em}-\frac{e^2}{2\, M_Z^2}X\, J_{em}^{KK}\cdot J_{em}^{KK}\, .
\end{eqnarray}
Using now Eqs.~(\ref{mWX}), (\ref{fermiun}) 
and (\ref{mZX}) one can relate
the angle $\theta$ to the Fermi constant as
\begin{equation}
\label{relacion1}
s^2_\theta\, c^2_\theta=\frac{\pi\alpha}{\sqrt{2}\, G_F\, M_Z^2}\,
\left(1+\Delta\right)\, ,
\end{equation}
where
\begin{equation}
\label{Delta}
\Delta=\left[\lL c_{2\alpha}c^2_\theta-s^4_\alpha s^2_\theta \right]\, X\, ,
\end{equation}
and the effective lagrangian can be cast as
\begin{eqnarray}
\label{lowneut}
\mathcal{L}^{neutral}_{low}&=&-\,4\,\frac{G_F}{\sqrt{2}}\, (1-\Delta)
\left\{
\left[J_Z-
s^2_\alpha
X\,  J_Z^{KK}\right]^2
-\,X\,J_Z^{KK}\cdot J_Z^{KK}\right\}\nonumber\\
&-&e A\cdot J^{em}-\frac{2\pi\alpha}{M_Z^2}X\, J_{em}^{KK}\cdot J_{em}^{KK}\, .
\end{eqnarray}

\section{The five-dimensional supersymmetric case}
\label{susy}

If the theory in 5D is supersymmetric, 
the field content must be extended to complete $N=2$ 
supermultiplets~\cite{peskin,alex}.
The on-shell field content of the gauge supermultiplet
is $\mathbb{V}=(V_\mu,V_5,\lambda^i,
\Sigma)$ where $\lambda^i\ (i=1,2)$ is a simplectic Majorana spinor 
and $\Sigma$ a real scalar in the adjoint representation;
$(V_\mu,\lambda^1)$ is  even under $\mathbb{Z}_2$ and 
$(V_5,\Sigma,\lambda^2)$  is odd. 
Matter and Higgs fields are arranged in $N=2$ hypermultiplets
that consist of  chiral and antichiral $N=1$ supermultiplets.
The chiral $N=1$ supermultiplets are even under  $\mathbb{Z}_2$ 
and contain  massless states. These will correspond to 
the SM fermions and Higgs.
Because of anomaly cancellation,
the Higgs doublet fields 
must come in pairs, 
$H_1$ and $H_2$.
For simplicity
we will just consider one pair of Higgs doublets.

Supersymmetry must be broken to give masses to all the superpartners.
We will not specify the way supersymmetry is broken,
but assume that the superpartner masses are of order $M_c$.
For a specific example see Ref.~\cite{alex}.
A priori one would think that integrating out the
superpartners at {\it tree level} does not  
lead to any effect in the SM lagrangian, as is the case for the 4D
supersymmetric extension of the SM.
Nevertheless, this is not true for the 5D supersymmetric case.
As we will show below, integrating out the 
scalar field $\Sigma$ induces a {\it tree-level} contribution
to the SM lagrangian.

The 5D lagrangian for the scalar $\Sigma$ can be written as
\begin{eqnarray}
\label{sigma5}
\mathcal{L}_{5D,\Sigma}={\rm Tr}\,\frac{1}{g_5^2}\,\left|D_\mu \Sigma\right|^2+
{\rm Tr}\,\frac{1}{g_5^2}\,\left|D_5 \Sigma\right|^2-\sum_{i=1}^2 
(1-\hi)H_i^\dagger\Sigma^2 H_i
\nonumber \\
-\left[\sum_{i=1}^2\hi H_i^\dagger (\partial_5 \Sigma) H_i +\frac{g_5^2}{2}
\sum_{\alpha}\left(\sum_{i=1}^2\hi H_i^\dagger T^\alpha H_i\right)^2 
\delta(x_5)\right]\delta(x_5)\, ,
\end{eqnarray}
where only the bosonic interactions of $\Sigma$ with the Higgs
have been considered. 
Using the fact that $\Sigma$ is odd under 
$\mathbb{Z}_2$-parity and integrating over $x_5$, we can obtain from 
Eq.~(\ref{sigma5}) a potential for the KK modes of $\Sigma$.
Due to the linear term in $\Sigma$ in Eq.~(\ref{sigma5}),
the VEV of $H_i$ on the boundary induces a VEV for the KK modes of $\Sigma$. 
Since the field $\Sigma$ transforms under SU(2)$_L\times$U(1)$_Y$ as 
$(\mathbf{3},0)+(\mathbf{1},0)$, its triplet-component VEV will give a
mass to the $W$ gauge boson, 
while its singlet-component does not couple to anything and is harmless.
Let us see this explicitly.
The triplet component of $\Sigma$ can be written, in matrix notation, as
\begin{equation}
\label{triplet}
\Sigma=\left(
\begin{array}{cc}
\xi_0/2 & \xi^+/\sqrt{2}\\  & \\                                  
\xi^-/\sqrt{2} &-\xi_0/2\\
\end{array}                          
\right)\, .
\end{equation}
Upon integration over $x_5$ and putting the Higgs fields $H_i$ at their 
VEVs, we obtain the potential for the KK modes of $\xi_0$
\begin{equation}
\label{potsig}
V_{\,\Sigma}=\frac{M_c^2}{2}\ \sum_{n=1}^\infty
\left\{n\,\xi_0^{(n)}-\frac{g}{\sqrt{2}}
\,\left(\htwo s_{\beta}^2-\hone c_{\beta}^2\right)\,
\frac{v^2}{M_c}
\right\}^2
+\frac{g^2}{8}\, c_{2\beta}^2\, v^4\ .
\end{equation}
Equation (\ref{potsig}) yields a VEV for $\xi_0^{(n)}$,
which is  given by
\begin{equation}
\label{xivev}
\langle\xi_0^{(n)}\rangle
=
\frac{1}{n}\,\left(\htwo s_{\beta}^2-\hone c_{\beta}^2\right)
 \,\frac{m_W}{M_c}\, v\, .
\end{equation}
Since these VEVs induce a mass term for the $W$, Eq.~(\ref{mWX}) must be 
corrected in the supersymmetric case to
\begin{equation}
\label{mWX2}
M_W^2=\left[1-\hone\htwo\, s_{2\beta}^2\, c^2_\theta X\right]\, m_W^2\, .
\end{equation}
Notice that the two terms, the one in Eq.~(\ref{mWX})
proportional to $-s^4_\alpha=-(\htwo s_{\beta}^2+\hone c_{\beta}^2)^2$ 
and the other proportional to
$(\htwo s_{\beta}^2-\hone c_{\beta}^2)^2$ from the triplet,
combine into an effective term proportional to $\hone\htwo\, s_{2\beta}^2$. 
This cancels in all cases, except if both Higgs doublets live on the boundary.
Because of the change in $M^2_W$, $\Delta$ defined in Eq.~(\ref{Delta})
must be replaced by
\begin{equation}
\label{Delta2}
\Delta=\left[\lL c_{2\alpha}\, c^2_\theta-s^4_\alpha
+\hone \htwo \, s^2_{2\beta}\, c^2_\theta\right]\, X\, .
\end{equation}
Therefore we conclude that for the 
5D supersymmetric case we obtain the same low-energy lagrangian as for 
the non-supersymmetric case, with the replacement of
Eqs.~(\ref{mWX}) and (\ref{Delta}) by Eqs.~(\ref{mWX2}) and (\ref{Delta2}),
respectively.

\section{Electroweak observables}
\label{electroweak}

The physical observables in the Standard Model can be predicted as functions
of some input parameters. One usually chooses as input parameters the best
measured ones, i.e. the Fermi constant $G_F=1.166\times
10^{-5}$ GeV$^{-2}$, the fine-structure constant $\alpha=1/137.036$ 
(or $\alpha(M_Z)=1/128.933$) and the
mass of the $Z$ gauge-boson $M_Z=91.1871$ GeV.
To make a reliable prediction of the other observables, one has to include
Standard Model radiative corrections as well as corrections due to the presence
of KK modes. In section~\ref{effective} we have only included tree-level
physics. 
This is correct for the KK modes since their masses will be constrained
to be large.
We will neglect electroweak
radiative corrections to $\mathcal{O}(X)$ effective operators, because their
contribution is inside the rest of uncertainties in the calculation, in 
particular those coming from the contribution of heavy 
($n\gtrsim 10$) KK modes, as was 
pointed out in section~\ref{introduction}. For practical purposes we use,
in the couplings of the KK modes, Eq.~(\ref{kkcouplings}), 
the weak angle~\cite{PDG} $s^2_\theta\equiv s^2_{M_Z}\simeq 0.231$.

The set of physical observables we have chosen for our fit is given in
Table 1, where the experimental values and Standard Model predictions used
in the fit are collected from Ref.~\cite{PDG}.
\begin{table}[H]
\centering
\begin{tabular}{||c|c|c||}\hline
Observable & Experimental value & Standard Model prediction \\ \hline
$M_W$ (GeV) & 80.394$\pm$0.042 & 80.377$\pm$0.023 ($-$0.036)\\
$\Gamma_{\ell\ell}$ (MeV) & 83.958$\pm$0.089 & 84.00$\pm$0.03 ($-$0.04)\\
$\Gamma_{had}$ (GeV)& 1.7439$\pm$0.0020 & 1.7433$\pm$0.0016 ($-$0.0005)\\
$A_{FB}^\ell$& 0.01701$\pm$0.00095 & 0.0162$\pm$0.0003 ($-0.0004$)\\
$Q_W$ & $-$72.06$\pm$0.46 & $-$73.12$\pm$0.06 ($+$0.01)\\
$\sum_{i=1}^{3}\left|V_{1i}\right|^2$ & 0.9969$\pm$0.0022 & 1 (unitarity)\\
\hline
\end{tabular}
\caption{Set of physical observables}
\label{tabla}
\end{table}
\noindent
The Standard Model predictions correspond to a Higgs mass $M_H=M_Z$ and a 
top-quark mass $m_t=173\pm 4$ GeV. 
The global shift in the prediction when $M_H$ is
shifted to 300 GeV is shown in parenthesis.
The observables in Table~\ref{tabla}
can be classified into LEP (high-energy) and
low-energy observables.

\subsection*{LEP observables}

The formalism we have developed in sections \ref{effective} and \ref{susy}
allows us to write 
the
prediction for these observables in terms of the Standard Model predictions.
In particular 
one can compute the
prediction for $M_W$ as
\begin{equation}
\label{MWpre}
M_W^2=\left(M_W^2\right)^{SM}\left[1-\frac{s^2_{\theta}}{c_{2\theta}}
\Delta+\Delta_W\,\right]\, ,
\end{equation}
where, for the non-supersymmetric case,
\begin{equation}
\Delta_W=s^2_{\theta}\,s^4_\alpha\, X\, ,
\label{delta1}
\end{equation}
whereas for the supersymmetric case
\begin{equation}
\Delta_W=\left[s^2_{\theta}\,s^4_\alpha+c^2_{\theta}\left(
\htwo s^2_\beta\,-\hone c^2_\beta\right)\right]\, X \ .
\label{delta2}
\end{equation}
We indicate with the $SM$ superindex the Standard Model prediction
including radiative corrections.

The rest of observables 
can be written 
in terms of effective vector and axial couplings of the $Z$,
which are defined as
\begin{equation}
\label{coup}
g_{V,A}^{\psi,\,eff}=\left(1-\frac{\Delta}{2}\right)
\left[ g_{V,A}^{\psi,\,SM}-s^2_\alpha X
\,g_{V,A}^{\psi,\,KK}\right]\, .
\end{equation}
In particular those selected in Table~\ref{tabla} are given
by
\begin{eqnarray}
\label{obspre}
{\displaystyle \frac{\Gamma_{\ell\,\ell}}{\Gamma_{\ell\,\ell}^{SM}} }&=&
\left[1-\Delta-2s^2_\alpha X\,
\frac{g_V^\ell\, g_V^{\ell,\,KK}+g_A^\ell\, g_A^{\ell,\,KK}}
{(g_V^{\ell})^2+(g_A^{\ell})^2}\right]\, ,\nonumber\\ && \nonumber\\
{\displaystyle
\frac{A_{FB}^\ell}{A_{FB}^{\ell,\,SM}} }&=&\left[1+2s^2_\alpha
X\,
\left(2\ \frac{g_V^\ell\, g_V^{\ell,\,KK}+g_A^\ell\, 
g_A^{\ell,\,KK}}
{(g_V^{\ell})^2+(g_A^{\ell})^2}-\frac{g_V^{\ell,\,KK}}{g_V^{\ell}}-  
\frac{g_A^{\ell,\,KK}}{g_A^{\ell}}  \right)  \right]\, ,\\&& \nonumber\\
{\displaystyle
\frac{\Gamma_{had}}{\Gamma_{had}^{SM}} }&=&\left[ 1-\Delta-2
s^2_\alpha
X\,
\frac{3\,g_V^d\, g_V^{d,\,KK}+
3\,g_A^d\, g_A^{d,\,KK}+2\,g_V^u\, g_V^{u,\,KK}+2\,g_A^u\, g_A^{u,\,KK}}
{3\,\left[(g_V^{d})^2+\,(g_A^{d})^2\right]+2\,\left[(g_V^{u})^2+\,(g_A^{u})^2
\right]} \right]\, ,\nonumber
\end{eqnarray}
%
In Eq.~(\ref{obspre}) we are assuming that the KK modes do not spoil
lepton and quark universality. The case with
different generations having different
KK couplings 
(which amounts to assuming that they  belong to different sectors,
either boundary or bulk, of the extra dimension) 
is completely obvious from the
previous expressions.

\subsection*{Low energy observables}

The low-energy observables are deduced from the expressions of the low-energy
lagrangians in the charged, Eq.~(\ref{lowcharged}), and neutral,
Eq.~(\ref{lowneutral}), sectors. In particular the observable $Q_W$ is obtained
from the low-energy lagrangian in the neutral sector, when selecting the
$(\bar{e}e)(\bar{u}u)$ and $(\bar{e}e)(\bar{d}d)$ crossed terms. 
Note that due 
to the (parity-non-conserving) 
way the electromagnetic current interacts with the KK modes of the photon,
there are also electromagnetic contributions to $Q_W$. Using the definition
of the effective couplings, the prediction for $Q_W$ is given by
\begin{equation}
\label{QW}
Q_W=(1-\Delta)\, Q_W^{SM}+16\, \delta\, Q_W\, ,
\end{equation}
where
\begin{eqnarray}
\label{deltaQW}
\delta Q_W&=&-\frac{1}{4}\, \frac{s^2_{\theta} \, c^2_{\theta}}
{c_{2\theta}}\, Z\, \Delta\nonumber\\
&+&\,X\, \left\{ (2Z+N)
\left[g_A^{e,\,KK} g_V^{u,\,KK}-\frac{1}{4}
s^2_\alpha
g_V^{u,\,KK}\right.\right.
\nonumber\\
&-&\left.\left.
s^2_\alpha
\left(\frac{1}{4}-\frac{2}{3}
s^2_{\theta}\right)g_A^{e,\,KK}+s^2_{\theta}c^2_{\theta}
g_{em,\,A}^{e,\,KK}\,g_{em,\,V}^{u,\,KK}\right]\right.\nonumber\\
&+&\left.(Z+2N)\left[g_A^{e,\,KK} g_V^{d,\,KK}-\frac{1}{4}
s^2_\alpha
g_V^{d,\,KK}\right.\right.
\nonumber\\
&-&\left.\left.
s^2_\alpha
\left(-\frac{1}{4}+\frac{1}{3}
s^2_{\theta}\right)g_A^{e,\,KK}+s^2_{\theta}c^2_{\theta}
g_{em,\,A}^{e,\,KK}\,g_{em,\,V}^{d,\,KK}\right]
\right\}\, ,
\end{eqnarray}
and the number of protons and neutrons in cesium is $Z=55$ and $N=78$.

The last observables we have selected in Table~\ref{tabla} are the quark mixing
angles $\left|V_{uq'}\right|$ where $q'=d,s,b$, 
which are subject to the 
unitarity condition
\begin{equation}
\label{unitar}
\sum_{q'=d,s,b}\left|V_{uq'}^{SM}\right|^2=1\, .
\end{equation}
The $\left|V_{uq'}\right|$ are extracted from quark $\beta$-decay amplitudes
$q'\rightarrow u\ell\bar{\nu}$ that can be described from the low-energy
lagrangian of Eq.~(\ref{lowcharged}). In fact the relevant piece of it
can be written as
\begin{eqnarray}
\label{CKMlag}
\mathcal{L}^{q'\,\beta-decay}=-\frac{G_F}{\sqrt{2}}\ \bar{\nu}_{\ell}
\gamma_\mu(1-\gamma_5) \ell\ \bar{u}\, V_{uq'}^{SM}\,\gamma^\mu (1-\gamma_5)\, 
q' \nonumber\\
\left\{1+c^2_{\theta}\,X\,\left[(\lL-\qL)
s^2_\alpha
+\lL\left(\qL-1\right)
 \right] \right\}+{\rm h.c.}\, .
\end{eqnarray}
In the presence of KK modes the relation (\ref{unitar}) yields
\begin{equation}
\label{unitar2}
\sum_{q'=d,s,b}\left|V_{uq'}\right|^2=1+2\,
c^2_{\theta}\,X\,\left[(\lL-\qL)
s^2_\alpha
+\lL\left(\qL-1\right) \right]\, ,
\end{equation}
which provides a further contribution to the fit after 
using the experimental value for
$\sum_{q'}\left|V_{uq'}\right|^2$ given in Table~\ref{tabla}. A further
(radiative) contribution to $\sum_{q'}\left|V_{uq'}\right|^2$ was
studied in Ref.~\cite{MS}, where the KK modes $Z^{(n)}$ were exchanged in
box diagrams. We will neglect this contribution since we are neglecting in 
our analysis radiative corrections involving KK modes. This  latter 
contribution is negligible with respect to the tree-level one (\ref{unitar2})
whenever the contribution of $W^{(n)}$ is non-zero, and in cases where it
vanishes (e.g. for $\lL=\qL$) it would be overwhelmed by 
the tree-level KK contribution to  other observables.

\section{Bounds from electroweak measurements}
\label{bounds}

In this section we will apply the results of section~\ref{electroweak}
to find bounds on the compactification scale $M_c$ for the different cases.
We will use the observables, computed in Eqs.~(\ref{MWpre}) to
(\ref{unitar2}), whose experimental values and SM predictions are 
listed in Table~\ref{tabla}. In all cases we will make a $\chi^2$ fit and
compute 95\% c.l. bounds. In particular we have computed the function 
$\chi^2(X)$ as
\begin{equation}
\chi^2(X)=\sum_j\frac{\left(\mathcal{O}_j(X)-\mathcal{O}_j^{\,exp}\right)^2}
{\left(\Delta\mathcal{O}_j\right)^2}\, ,
\label{chi}
\end{equation}
where the sum is extended to the used observables $\mathcal{O}_j$ of 
Table~\ref{tabla} and the lower bound on $M_c$ is computed as the 
solution to $\chi^2(X)=\chi^2_{min}+1.96^2$, where $\chi^2_{min}$ is the
minimum of $\chi^2(X)$, provided that $\chi^2_{min}$ corresponds to a point
$X>0$. Otherwise we have followed the prescription of Ref.~\cite{Feldman}.

All the results will depend on the ratio between
the VEV of $H_2$ and $H_1$, $\tan\beta$, which also measures the effective 
mixing angle between the VEV on the brane and  in the bulk  
--- see Eq.~(\ref{alpha}). 
This effective angle also depends on the location of the
Higgs fields $H_{1,2}$, i.e. on whether they are  localized 
on the 4D brane or
spread out in the 5D bulk. All possible situations are 
taken into account by
the parameters $\hi=0,1$, which give rise to four distinct possibilities.

Concerning fermion fields they can also be either localized on the brane or
in the bulk. All possibilities are encompassed by the parameters,
$\qLi,\,\uRi,\,\dRi,\,\lLi,\,\eRi=0,1$, which yield a very large number of different
possibilities. We will reduce the number of cases:
\begin{itemize}
\item
First, by assuming universality of different generations. 
In practice this means
that: $\qLi\equiv\qL$, $\uRi\equiv\uR$, $\dRi\equiv\dR$, $\lLi\equiv\lL$ and
$\eRi\equiv\eR$. Giving up universality will be done in 
section~\ref{flavor}.
\item
Then, by imposing the different Yukawa couplings responsible 
for fermion masses to be compatible with the $\mathbb{Z}_2$ orbifold 
action~\footnote{In the language of the heterotic string, localized states on
the brane correspond to twisted states ($T$) and states in the bulk to 
untwisted ($U$) ones. 
Invariance under the orbifold group selects the Yukawa couplings
of the type $TTU$ and $UUU$. The latter are expected to be 
suppressed upon compactification of the large dimension, 
by an extra factor of $1/(M_s R)$ and will not be considered in our 
subsequent analysis.}. 
\end{itemize}
In particular the latter condition selects unambiguously a number of cases 
from the very existence of Yukawa couplings: 
$H_2 \bar{q}_L u_R,\, H_1 \bar{q}_L d_R,\, H_1 \bar{\ell}_L e_R$. 
In this case the $\varepsilon$-parameters should satisfy the set of equations
\begin{eqnarray}
\htwo+\qL+\uR&=&2\, , \nonumber\\
\hone+\qL+\dR&=&2\, , \nonumber\\
\hone+\lL+\eR&=&2\, . \quad 
\label{casos}
\end{eqnarray}
The cases consistent with Eq.~(\ref{casos}) are, for the two Higgs fields 
localized on the brane:
\begin{equation}
\hone=\htwo=1\ \left\{
\begin{array}{l}
\lL=\uR=\dR=1 \\
\uR=\dR=\eR=1 \\
\qL=\eR=1\\
\qL=\lL=1
\end{array}\right.\, .
\label{todos}
\end{equation}
If at least one of the Higgs fields is living in the 
bulk the relevant cases are:
\begin{eqnarray}
\hone=1,\ \htwo=0\ && \left\{
\begin{array}{l}
\qL=\lL=\uR=1 \\
\qL=\uR=\eR=1 
\end{array}\right.\, ,
\nonumber\\
\hone=0,\, \htwo=1&&\ \quad \qL=\lL=\dR=\eR=1\, ,\nonumber\\
\hone=\htwo=0&&\ \quad \qL=\lL=\uR=\dR=\eR=1\, .
\label{notodos}
\end{eqnarray}
In all cases the unspecified values of the $\varepsilon$-parameters are 
supposed to be zero. Also the cases of the 5D extension of the 
(non-supersymmetric) SM considered in section~\ref{effective} and 
of the extension of the supersymmetric theory, section~\ref{susy}, should be
considered separately, since we know from the analysis in section~\ref{susy}
that  they give rise to different  KK effects.

\begin{figure}[H]
\centering
\epsfig{file=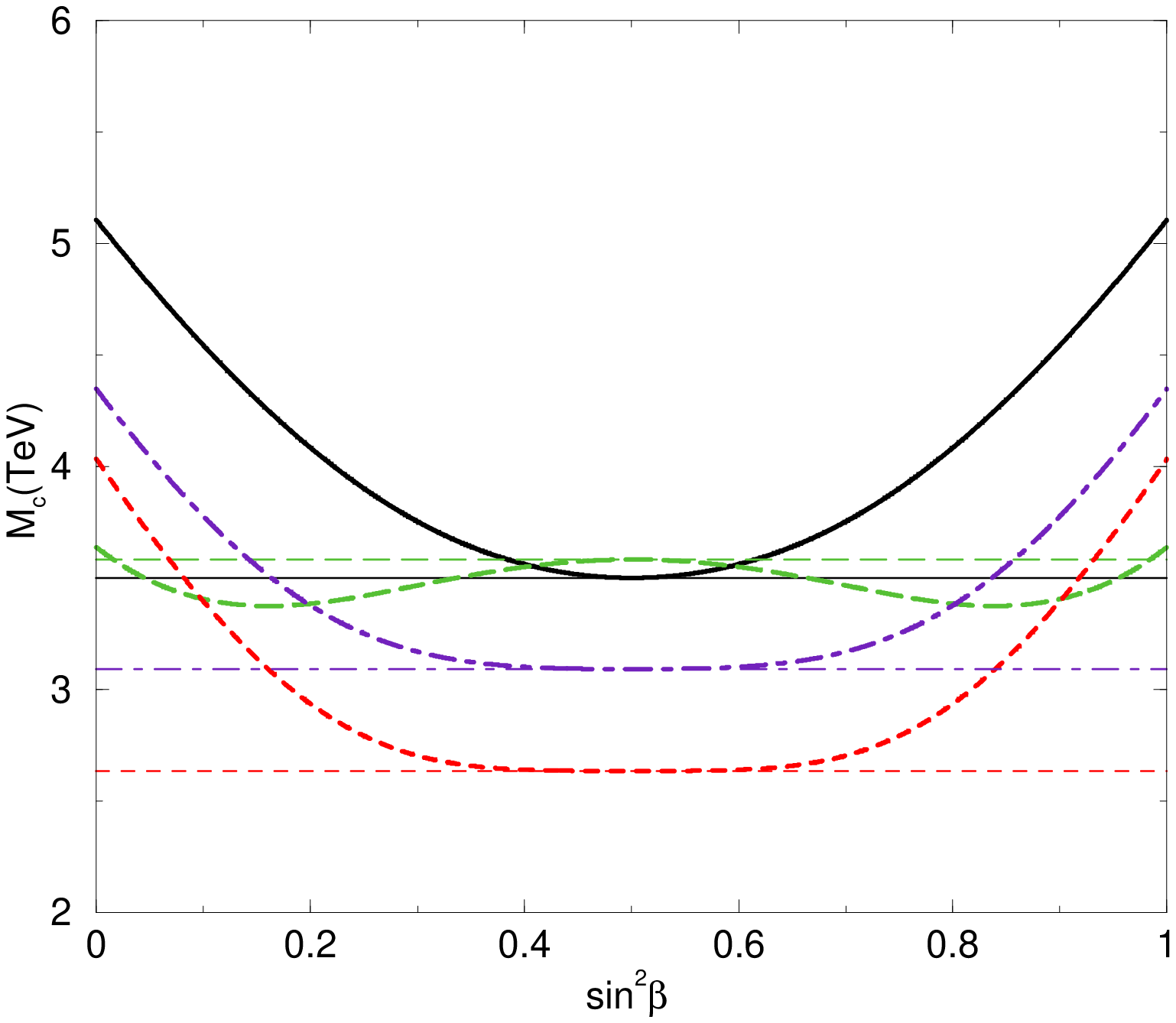,width=.5\linewidth}
\caption{Lower bounds on $M_c$ corresponding to the case $\hone=\htwo=1$ and
$\lL=\uR=\dR=1$ (solid line), $\qL=\eR=1$ (long-dashed), 
$\uR=\dR=\eR=1$ (short-dashed) and $\qL=\lL=1$ (dash-dotted). Straight 
(non-straight) lines correspond to SM (MSSM) 5D extensions.}
\label{fig1}
\end{figure}

\begin{figure}[H]
\centering
\epsfig{file=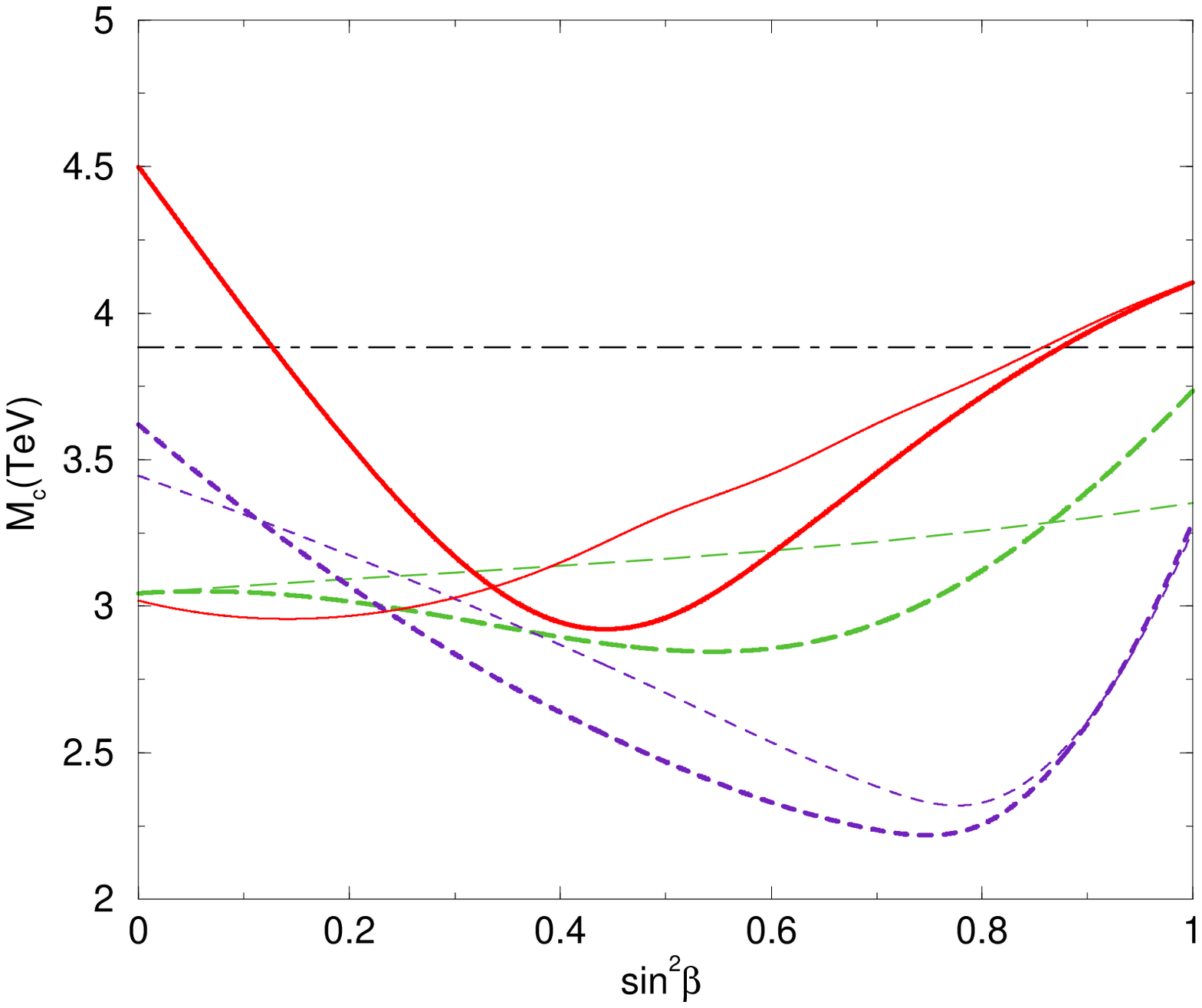,width=.5\linewidth}
\caption{Lower bounds on $M_c$ corresponding to the case
$\hone=1,\ \htwo=0$, $\qL=\lL=\uR=1$ (solid),
$\hone=1,\, \htwo=0,\, \qL=\uR=\eR=1$ (short-dashed), 
$\hone=0,\,\htwo=1,\,\qL=\lL=\dR=\eR=1$ (long-dashed) and 
$\hone=\htwo=0,\,\qL=\lL=\uR=\dR=\eR=1$ (dash-dotted). Thin (thick)
lines correspond to SM (MSSM) 5D extensions.}
\label{fig2}
\end{figure}

Our numerical results are summarized in  Figs.~\ref{fig1} and \ref{fig2},
which correspond to the cases considered in Eqs.~(\ref{todos}) and 
(\ref{notodos}) respectively.


\begin{figure}[H]
\centering
\epsfig{file=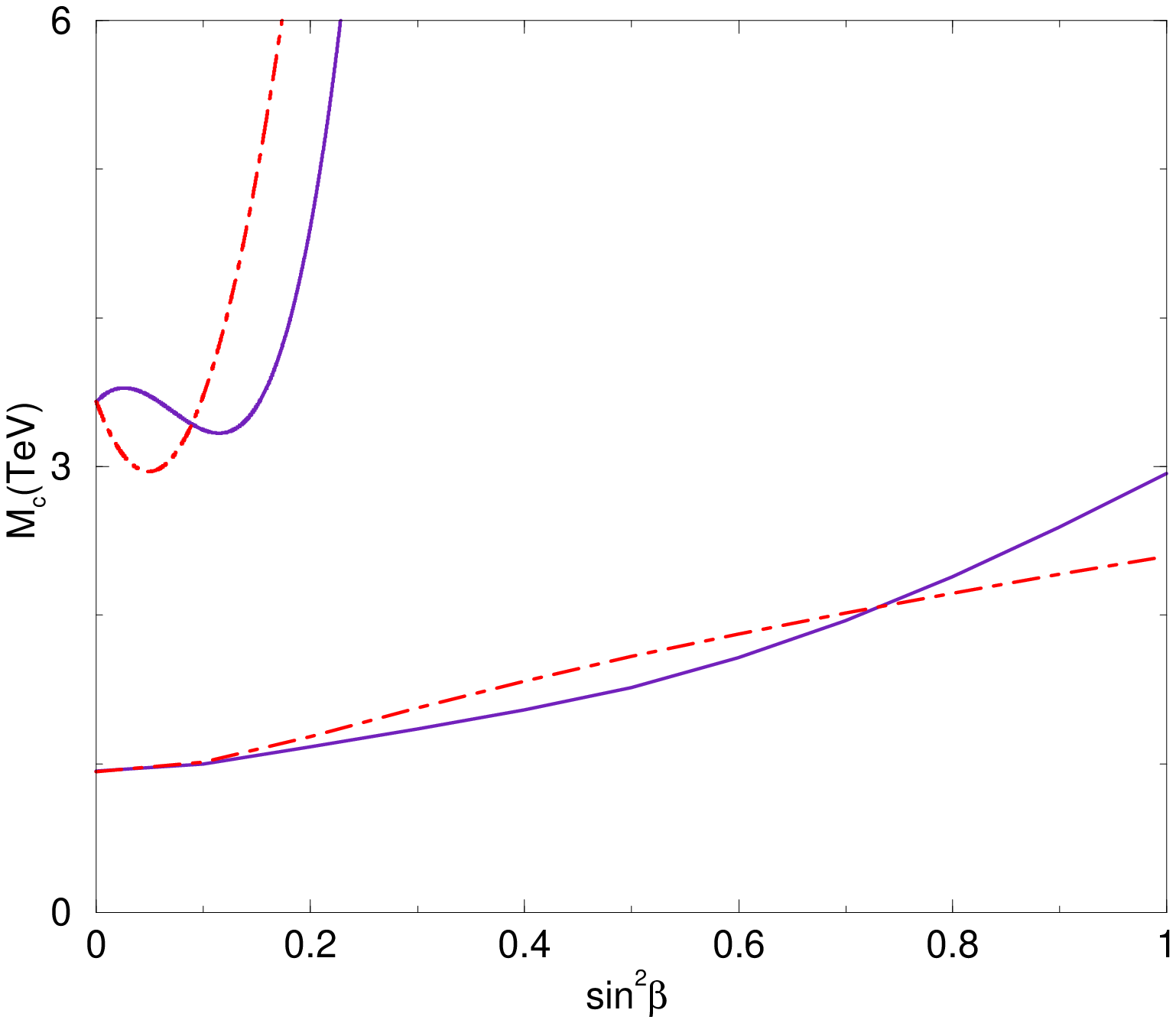,width=.5\linewidth}
\caption{Lower (and upper) 95\% c.l. bounds on $M_c$, as functions of
$s^2_\beta$, for the case $\htwo=\eR=\dR=1$, and $\hone=\uR=\lL=\qL=0$.
Solid (dash-dotted) lines correspond to the MSSM (SM) case.}
\label{fig3}
\end{figure}

We could also relax
the condition (\ref{casos})
and consider 
more general cases. They generically lead 
to bounds on
$M_c$ of order a few TeV, as those in Figs.~\ref{fig1} and \ref{fig2}. 
However,
some special cases can be constructed {\it \`a la carte} where the fit to the
electroweak observables is particularly better than in the SM. To this end
we can realize that the SM prediction for all observables in Table~\ref{tabla}
lies close enough to the corresponding experimental value, except for 
the case of 
$Q_W$(Cs) whose SM prediction falls more than 2$\sigma$ away from the
experimental result. Then new physics with $\delta\,Q_W>0$, and not touching
significantly the rest of observables, is required. A quick glance at
Eqs.~(\ref{Delta2}), (\ref{coup}), (\ref{deltaQW}) and (\ref{unitar2}) 
shows that
this is indeed the case when $e_R$ and $d_R$ are on the brane, 
$\eR=\dR=1$, while all other fields, $H_{1,2}$, $u_R$, $q_L$ and 
$\ell_L$ live in
the 5D bulk. In this case all observables in Table~\ref{tabla} are unmodified,
except $Q_W$, which experiences a positive shift,
\begin{equation}
\delta\, Q_W=\frac{1}{12}\,(Z+2\,N)\,s^2_{\theta}\ X\, .
\label{QWespe}
\end{equation}
Now the fit to the observables of Table~\ref{tabla} has, with respect to the
SM fit, $\Delta\chi^2\simeq -5.3$ and $\chi^2_{min}$ corresponds, using
Eq.~(\ref{QWespe}), to $M_c=1.30$ TeV, while the 95\% c.l. upper and lower
bounds are $0.95$ TeV $\leq M_c\leq 3.44$ TeV. Actually a model with these
qualitative features is not unique. Even if one of the Higgs fields, e.g.
$H_2$, lives in the bulk the contribution to $\delta\, Q_W$ can be positive
and provide low values of the 95\% c.l. lower (and even upper) bound. These
cases are exemplified in Fig.~\ref{fig3}, where the bounds 
are shown for the case
in which only the fields $H_2$, $e_R$ and $d_R$ live on the boundary while the
rest of the fields live in the bulk.

\section{Flavor and CP-violating physics}
\label{flavor}

Up to now we have considered universality of KK interactions with respect to
the different families of SM fermions. In this section we will
consider the case where the different families 
of quarks and leptons are localized in different points of the
extra dimension. This possibility
has been motivated in order to explain the difference 
of masses of the families \cite{as}.
Also certain models from branes seem to 
lead to two families living in the bulk and one living on the boundary
\cite{st}.
In all these situations one has flavor-violating 
interactions and therefore stronger bounds on $M_c$
coming from the experimental constraints on flavor  
and CP-violating physics.
It is easy to understand how these flavor-violating interactions arise.
The coupling of the gauge KK-excitations to the SM fermions 
is proportional to
$\varepsilon^{\psi_i} \cos (nx_5/R)$,
where $\varepsilon^{\psi_i}=1\ (0)$ for localized (not localized) 
fermions. Therefore if fermions live at 
different points of the extra dimension (or they have different 
$\varepsilon^{\psi_i}$), they will couple differently to the 
KK excitations, leading to flavor-violating couplings.

Let us consider in detail the case for the first two families. 
In the interaction basis, the flavor-violating couplings are
given by (we only consider the KK excitations of the gluons 
$G^{A\, (n)}_\mu$ 
since they should provide the strongest constraint):
\begin{eqnarray}
-{\cal L}&=&(\bar d_R\ \bar s_R) V^{d\, \dagger}_R 
\left(
\begin{array}{cc}
m_d & 0\\0 & m_s
\end{array}
\right)
 V^d_L
\left(
\begin{array}{c}
d_L\\ s_L
\end{array}
\right)
+
\frac{g}{\sqrt{2}}W_\mu(\bar u_L\ \bar c_L)\gamma^\mu 
\left(
\begin{array}{c}
d_L\\ s_L
\end{array}
\right)
\nonumber\\
&+&\sum_{n=1}^\infty
\Bigg[\sqrt{2}g_s G^{A\, (n)}_\mu 
(\bar d_L\ \bar s_L) \gamma^\mu T^A
\left(
\begin{array}{cc}
c_1^{(n)}& 0\\0 & c_2^{(n)}
\end{array}
\right)
\left(
\begin{array}{c}
d_L\\ s_L
\end{array}
\right)
+(L\rightarrow R)
\Bigg]\nonumber\\
&+&(d,s\leftrightarrow u,c)\, ,
\label{lagra1}
\end{eqnarray}
where $V^d_{L,R}$ are generic unitary matrices, and we have defined
$c_{1,2}^{(n)}\equiv \varepsilon^{\psi_{1,2}}\cos\left(nx^{(1,2)}_5/R\right)$ 
where $x^{(1,2)}_5$ is the position, along the fifth dimension, of the 
left-handed quarks of the first and second family respectively. 
Going to the mass-eigenstate basis, we have
\begin{eqnarray}
-{\cal L}&=&(\bar d_R\ \bar s_R) 
\left(
\begin{array}{cc}
m_d& 0\\ 0&m_s
\end{array}
\right)
\left(
\begin{array}{c}
d_L\\ s_L
\end{array}
\right)
+
\frac{g}{\sqrt{2}}W_\mu(\bar u_L\ \bar c_L)\gamma^\mu V^u_L V^{d\, \dagger}_L
\left(
\begin{array}{c}
d_L\\ s_L
\end{array}
\right)
\nonumber\\
&+&\sum_{n=1}^\infty
\Bigg[\sqrt{2}g_s G^{A\, (n)}_\mu 
(\bar d_L\ \bar s_L) \gamma^\mu T^A V^d_L
\left(
\begin{array}{cc}
c_1^{(n)}& 0\\ 0&c_2^{(n)}
\end{array}
\right)
V_L^{d\, \dagger}
\left(
\begin{array}{c}
d_L\\ s_L
\end{array}
\right)
+(L\rightarrow R)
\Bigg]\nonumber\\
&+&(d,s\leftrightarrow u,c)\, ,
\label{lagra2}
\end{eqnarray}
showing two sources of flavor violation, one from the CKM matrix 
$V^u_L V^{d\, \dagger}_L$ and another (if $c_1^{(n)}\not=c_2^{(n)}$) from
\begin{equation}
 U^d_L\equiv  V^d_L
\left(
\begin{array}{cc}
c_1^{(n)}& 0\\0 &  c_2^{(n)}
\end{array}
\right)
V_L^{d\, \dagger}\, ,
\label{U}
\end{equation}
mediated by the KK gluons. Notice that for general $V^{u,d}_{L,R}$ the
phases cannot be rotated away by field redefinitions
and we can have CP violation with only two families.

The flavor-violating couplings of the KK gluons
mediate flavor-changing neutral currents (FCNC) at tree level.
For the down-type quarks
the effective $\Delta S=2$ 
lagrangian mediated by the KK gluons
is given by
\begin{equation}
{\cal L}^{\Delta S=2}=\sum_{n=1}^\infty \frac{2g_s^2}{3n^2M^2_c}  
\left[U^d_{L\{12\}} 
\bar d_L \gamma^\mu s_L+(L\rightarrow R)+{\rm h.c.}\right]^2\, ,
\label{fcnc}
\end{equation}
where by $U^d_{L\{ij\}} $ we denote the $\{ij\}$ element of the matrix 
$U^d_L$.  From Eq.~(\ref{U}), we have
using the unitarity of $V^d_L$
\begin{equation}
\sum_{n=1}^\infty \frac{U^{d\, 2}_{L\{12\}}}{n^2}=
\left(V^{d\,* }_{L\{21\}}V^{d\, }_{L\{11\}}\right)^2\sum_{n=1}^\infty 
\left(\frac{c_1^{(n)}-c_2^{(n)}}{n}\right)^2
\label{mixing}
\, ,
\end{equation}
where the sum over the KK excitations is given by
\begin{equation}
\sum_{n=1}^\infty
\left(\frac{c_1^{(n)}-c_2^{(n)}}{n}\right)^2
=
\left\{\begin{array}{ll}
{\displaystyle 
\frac{\pi}{2\,R}\left|x^{(1)}_5-x^{(2)}_5 \right| }\, \ \ \
&{\rm for}\ \varepsilon^{\psi_{1,2}}\not=0\\ &\\
{\displaystyle 
\frac{\pi^2}{24} + \frac{1}{2}\left(\frac{x^{(2)}_5}{R}-
\frac{\pi}{2}\right)^2 }\, \ \ \ 
&{\rm for}\ \varepsilon^{\psi_{1}}=0\, ,\varepsilon^{\psi_{2}}\not=0
\end{array}\right.\, .
\label{sum}
\end{equation}
Notice that the sum over $n$ in Eq.~(\ref{sum}) is extended from $n=1$ 
to $n=\infty$, as was done in Eq.~(\ref{equis}), although, properly speaking,
it should have been extended up to $n\sim \Lambda\, R$,
where $\Lambda$ is an effective 
ultraviolet cutoff related to the tension of the 4D brane.
An equivalent
way the cutoff appears is through the coupling of the
$n$-KK to the brane fields, $g_n$, which can
be computed and leads to an exponential drop-off as \cite{ignatios,KKc}
\begin{equation}
g_n^2 = g^2 e^{-\frac{n^2}{\Lambda^2 R^2}}\, .
\label{drop}
\end{equation}
The value of the cutoff $\Lambda$ has been studied in string theory, and in the
effective theory of the brane world. 
It is found  that $\Lambda=c\, \mu$, where $\mu$ is the tension of the brane
and  $c$ is a model-dependent coefficient \cite{ignatios,KKc}.
The sum in (\ref{equis}) and in (\ref{sum}) for 
$\varepsilon^{\psi_{1}}=0\, ,\varepsilon^{\psi_{2}}\not=0$ is dominated
by the contribution of the first $\sim$10 KK,  
and thus is not sensitive to the
value of $\Lambda$. 
Equation (\ref{sum}) for $\varepsilon^{\psi_{1,2}}\not=0$
is also insensitive to the value of $\Lambda$, but only for large distances,
$|x_5^{(1)}-x_5^{(2)}|\, \Lambda \gtrsim 1$.
However, for small distances
\footnote{
If $\Lambda$ is close to the cutoff scale of the 5D field theory ($M_{s}$),
calculations  at distances smaller than $\Lambda$ will only
make sense in  the underlying theory  (strings). 
},
$|x_5^{(1)}-x_5^{(2)}|\, \Lambda \lesssim 1$, and for values of
the center-of-mass of the two branes $(x_5^{(1)}+x_5^{(2)})/2R=\mathcal{O}(1)$,
the expression (\ref{sum}) should be replaced by
\footnote{If the two branes are close to the boundaries of the 
orbifold, $(x_5^{(1)}+x_5^{(2)})/2R\simeq 0,\pi$,
the sum is given by
$\sum_n 
(c_1^{(n)}-c_2^{(n)})^2/n^2
\simeq\Lambda^3(x^{(1)}_5-x^{(2)}_5)^4/12R$.}
\begin{equation}
\sum_{n} 
\left(\frac{c_1^{(n)}-c_2^{(n)}}{n}\right)^2
\simeq\frac{\Lambda}{2\,R} (x^{(1)}_5-x^{(2)}_5)^2\ .
\label{sumsmall}
\end{equation}

Armed with the  lagrangian (\ref{fcnc}), we can calculate the contribution
to $\Delta m_K$ and $\varepsilon_K$,
\begin{eqnarray}
\Delta m_K&=&\frac{{\rm Re}\langle K| -{\cal L}^{\Delta S=2} |\bar K\rangle}
{m_K}
=3.5\times 10^{-15} \ {\rm GeV}\, ,\\  
\left|\varepsilon_K\right|
&=&\frac{\left|{\rm Im} \langle K| -{\cal L}^{\Delta S=2}|\bar K\rangle\right|}
{2\sqrt{2}m_K\Delta m_K}
=2.3\times 10^{-3}\, .
\end{eqnarray}
Let us just consider the left-handed KK contribution 
to $\Delta m_K$ and $\varepsilon_K$. Requiring these to be smaller than the 
experimental value, we obtain respectively 
(using the vacuum-insertion approximation)
\begin{eqnarray}
M_c&\gtrsim&400 \ {\rm TeV}
\left(\frac{\sqrt{\sum_n {\rm Re}U^{d\, 2}_{L\{12\}}/n^2}}{0.3}\right)\, ,
\label{condm}
\\
M_c&\gtrsim&5000 \ {\rm TeV}
\left(\frac{\sqrt{\sum_n {\rm Im}U^{d\, 2}_{L\{12\}}/n^2}}{0.3}\right)\, ,
\label{conde}
\end{eqnarray}
where for the numerical bound we have assumed that the mixing angles
$V^d_{L\{ij\}}$  are similar to those in the CKM matrix
and that one family lives on the boundary ($c_1^{(n)}=1$) and the other 
in the bulk ($c_2^{(n)}=0$) so that Eq.~(\ref{sum}) can be applied.

The constraints on $M_c$ for
other scenarios 
can be easily read off from Eqs.~(\ref{condm}) and (\ref{conde}).
For example, let us consider the case in which the first and second families
are localized at different points of the extra dimension.
This could nicely explain the smallness of their masses, since
mass terms can only arise from exponentially small overlaps of the
wave functions of the fields \cite{as}:
\begin{equation} 
M_{d\ \{ij\}}\simeq e^{-\mu^2|\bar x^{(i)}_5-x^{(j)}_5|^2/2}\ v \, ,
\label{massm}
\end{equation}
where  $\bar x^{(i)}_5$ refers to the position in 
 the extra dimension of the
right-handed $i$ quark 
and  $x^{(j)}_5$  to the position of the left-handed $j$ quark.
Requiring that $M_{d\ \{22\}}= m_s$ and
$M_{d\ \{21\}}=m_s s_{\theta_C}$ ($s_{\theta_C}\simeq 0.22$ is the
Cabibbo angle), we have that the distance between the $s_L$ and $d_L$ is
\begin{equation} 
|x^{(1)}_5-x^{(2)}_5|\simeq \frac{\sqrt{2}}{\mu}
\left[
\ln^{1/2}\left(\frac{v}{m_s s_{\theta_C}}\right)
-\ln^{1/2}\left(\frac{v}{m_s}\right)
\right]\simeq 0.4\mu^{-1}\, .
\end{equation}
Using Eq.~(\ref{sumsmall}), we have from
Eqs.~(\ref{condm}) and (\ref{conde}) 
respectively the bounds 
\begin{eqnarray}
M_c &\gtrsim& 25\ {\rm TeV} \sqrt{10\Lambda/R\mu^2}\, ,\\
M_c &\gtrsim& 300\ {\rm TeV} \sqrt{10\Lambda/R\mu^2}\, ,
\end{eqnarray}
where we have normalized the bound to the case $\mu\sim\Lambda$ and 
$\Lambda R\sim 10$.
These bounds are quite strong and disfavor this type of scenarios 
in models with TeV-string scale. On the same footing, 
we can also get constraints from the up-type quark sector. Since 
$(\Delta m_D/m_D)_{exp}\simeq 10\,(\Delta m_K/m_K)_{exp}$, we get
a constraint that is a factor of $\sim 0.3$ weaker than that in
Eq.~(\ref{condm}).

The KK excitations of the gluon
can also induce $\Delta S=1$ terms in the 
lagrangian that contribute to $\varepsilon^\prime_K/\varepsilon_K$. 
We find that the dominant contribution is given by
\begin{eqnarray}
\left|\frac{\varepsilon^\prime_K}{\varepsilon_K}\right|&\simeq& 
\frac{\omega}{\sqrt{2}\left|\varepsilon_K\right| {\rm Re}A_0}
\sum_n \frac{g^2_s}
{n^2M^2_c}
{\rm Im}\Bigg\{\frac{U^d_{L\, \{12\}}}{12}
\Big[(U^u_{R\,\{11\}}+2U^d_{R\,\{11\}})\, \langle 
(\pi\pi)_{I=0}|Q_6|K\rangle\nonumber\\
&-&
\frac{2}{\omega}
(U^u_{R\,\{11\}}-U^d_{R\,\{11\}})\, 
\langle (\pi\pi)_{I=2}|Q_8|K\rangle\Big]\Bigg\}\, ,
\label{epp}
\end{eqnarray}
where we follow the notation of Ref.~\cite{buras}.
Note that, since $\omega\simeq 1/22$, 
the second term of Eq.~(\ref{epp}) gives the dominant contribution
if there is isospin breaking in the right-handed sector
(i.e. $U^u_R\not= U^d_R$). 
This is the case whenever the $u_R$ and the $d_R$ live in different
points of the extra dimension. 
Considering this latter case, e.g. $U^d_{R\{11\}} \gg U^u_{R\{11\}}$, 
we obtain  from Eq.~(\ref{epp}) and 
$\left|\varepsilon^\prime_K/\varepsilon_K\right|_{exp}<2.5\times 10^{-3}$:
\begin{equation}
M_c\gtrsim 150 \ {\rm TeV}
\left(\frac{\sqrt{
\sum_n {\rm Im}[U^d_{L\,\{12\}}U^d_{R\,\{11\}}]/n^2}}{0.6}\right)\, .
\label{condep}
\end{equation}
This bound is not competitive with that from $\varepsilon_K$ 
[Eq.~(\ref{conde})]
if the mixing angles in $U^d_L$ are 
of order $s_{\theta_C}$.
Nevertheless, since the bound from Eq.~(\ref{condep}) scales as the
square-root of the mixing angle, instead of linearly as that in
Eq.~(\ref{conde}), we have that for 
small mixings, i.e. Im$\{U^d_{L\, \{12\}}\}
\lesssim 2\times 10^{-4}$,
the bound (\ref{condep}) becomes the strongest one.
In other words, sizeable contributions to both 
$\varepsilon^\prime_K/\varepsilon_K$ and $\varepsilon_K$
from KK gluons occurs for Im$\{U^d_{L\, \{12\}}\}
\sim 2\times 10^{-4}$ and $M_c\sim 4$ TeV.

For models where only the third family lives in another point of the 
space ($c_1^{(n)}=c_2^{(n)}\not=c_3^{(n)}$), the above bounds  also apply,
with the only difference that now 
Eq.~(\ref{mixing}) must incorporate the third family. One obtains
\begin{equation}
\sum_{n=1}^\infty \frac{U^{d\, 2}_{L\{12\}}}{n^2}=
\left(V^{d\,*}_{L\{23\}}V^{d}_{L\{13\}}\right)^2\sum_{n=1}^\infty 
\left(\frac{c_1^{(n)}-c_3^{(n)}}{n}\right)^2
\label{mixing3}
\, .
\end{equation}
For mixing angles similar to those in the CKM matrix
(Im$\{V^{d}_{L\{23\}}V^{d}_{L\{13\}}\}\sim 2\times 10^{-4}$)
we get from Eq.~(\ref{condep}) 
\begin{equation}
 M_c\gtrsim 4\ {\rm TeV}\, .
\end{equation}
It is interesting to remark that in this case the contribution to 
$\varepsilon_K$ and $\varepsilon^\prime_K/\varepsilon_K$  
are both saturated for similar values of the compactification scale,
$M_c\sim 4$ TeV. $B$ physics can also put constraints on $M_c$.
{}From the experimental value of $\Delta m_B$,
we obtain
\begin{equation}
M_c\gtrsim 2 \ {\rm TeV}
\left(\frac{\sqrt{\sum_n {\rm Re}\, U^{d\, 2}_{L\{13\}}/n^2}}{0.005}\right)\, ,
\label{condm2}
\end{equation}
where again we have considered that the mixing angles between the first and 
third families are similar to those in the CKM matrix.

Let us finally consider the lepton sector. 
Bounds on family-violating couplings can be 
obtained, for example, from the experimental upper bound
on BR$(\mu\rightarrow 3e)$.
If the $\mu$ and $e$ live in different points,
there will be a family transition $\mu-e$
mediated by the KK excitations of the $W_3$. We obtain
\begin{equation}
{\rm BR}(\mu\rightarrow 3e)=\left|\sum_n
\frac{\sqrt{2}m^2_W} 
{n^2M_c^2} U^\ell_{L\{21\}}U^\ell_{L\{11\}}
\right|^2\, ,
\end{equation}
where now the mixing angles refer to the ones in the lepton sector.
From BR$(\mu\rightarrow 3e)_{exp}<10^{-12}$ we obtain the bound
\begin{equation}
M_c\gtrsim 30 \ {\rm TeV}
\left(\frac{\sqrt{\sum_n |U^\ell_{L\{21\}}U^\ell_{L\{11\}}|/n^2}}{0.3}\right)
\, ,
\end{equation}
where for the numerical estimate we assumed $c_1^{(n)}=1$ and $c_2^{(n)}=0$ 
and $|V^\ell_{L\{11\}}|\simeq 1$, $|V^\ell_{L\{21\}} 
|\sim\sqrt{m_e/m_\mu}\simeq 0.07$.

\section{Conclusion}
\label{conclusion}

In this paper we have studied the implications of a TeV$^{-1}$ size
dimension, where SM gauge-bosons propagate, on electroweak
and flavor physics. This work generalizes those existing in the literature
in two aspects:
\begin{itemize}
\item
We have allowed for the possibility that the different fermions 
 (as well as
Higgs fields) 
live  in the bulk of the
extra dimension or  are localized at different points of it, 
whereas previous
analyses had only considered the case where all SM fermions are stuck at
the boundary of the extra dimension.
\item
We have analyzed   the extension of the SM to
five dimensions and
its minimal supersymmetric generalization.
We have found interesting tree-level  effects associated to the presence of
supersymmetric partners. Previous analyses in the literature only considered
the case of the 5D SM extensions.

\end{itemize}
Concerning the possibility of different 
locations for the different fermions,
 we
have considered
a wide variety of cases. 
Assuming 
universality in family space, 
we have seen that the 
lower bound on $M_c$
 is very model-dependent, but it is generically around
$2$--$5$ TeV. Nevertheless, we have found
particular models
where the correction to the weak charge $Q_W$
is positive, as
 required by the experimental data,
whereas the other observables are unchanged. 
These cases provide
 a global fit
to the electroweak observables  better than that  of the  SM.
The model with the best fit yields
the 95\% c.l. region: 
$0.95\, {\rm TeV}\lesssim\, M_c \lesssim 3.44\, {\rm TeV}$.

In the 5D supersymmetric case,
we have found that one of the 
supersymmetric partners is 
a scalar SU(2)$_L$-triplet
that acquires a  VEV of  $\mathcal{O}(R M_Z)$ whenever a
Higgs lives on the 4D boundary.
This VEV modifies the SM relation between the $Z$ and $W$ masses
and consequently the analysis of the global fit to the electroweak data.
We find very different bounds on $M_c$ from the 
non-supersymmetric case.

Giving up lepton and quark universality and allowing the 
different families to be located at different points of the 
fifth dimension, 
we  found that the KK excitations generate dangerous FCNC
at the tree level.
The effect on flavor observables 
such as $\Delta m_K$, $\varepsilon_K$ and 
$\varepsilon^\prime_K$, provides very stringent limits
on $M_c$,  as strong as $\sim 5000$ TeV.
This seems to disfavor this type of scenarios in the context of TeV-strings.

We want to conclude  by stressing some of
the importance of the analysis carried out
here.
If indirect effects 
of an extra dimension (such as those considered here) 
put already strong bounds on $M_c$, 
it will make very unlikely the 
direct detection of the  KK excitations of a SM field 
in future colliders \cite{KKpro,rw}.
This would be the real test of an extra dimension.
Combining 
the analysis here with that in Ref.~\cite{KKpro,rw},
one learns that only the LHC, that will probe KK excitations
up to $6$--$7$ TeV,
has a chance to discover an extra dimension. 
If the extra dimension treats families in a non-universal way,
the bounds found here put the size of the extra dimension far 
from the 
experimental reach.

\section*{Acknowledgements} 
We thank K. Benakli and G. Kane
 for useful discussions. The work of AD was supported 
by the Spanish Education Office (MEC) under an~\emph{FPI} scholarship.   

\begin{thebibliography}{99}
%
\bibitem{ignatios} I.~Antoniadis, \PLB{246}{90}{377}; I.~Antoniadis,
C.~Mu\~noz and M.~Quir\'os, \NPB{397}{93}{515}; I.~Antoniadis, K.~Benakli and
M.~Quir\'os, \PLB{331}{94}{313}; I.~Antoniadis and K.~Benakli, 
\PLB{326}{94}{69}; K.~Benakli, \PLB{386}{96}{106}; I.~Antoniadis and
M.~Quir\'os, \PLB{392}{97}{61};
E.~C\'aceres, V.S.~Kaplunovsky and I.M.~Mandelberg,
\NPB{493}{97}{73}.
%
\bibitem{lykken} J.D.~Lykken, \PRD{54}{96}{3693}.
%
\bibitem{add} N.~Arkani-Hamed, S.~Dimopoulos and
G.~Dvali, \PLB{429}{98}{263};
I.~Antoniadis, N.~Arkani-Hamed, S.~Dimopoulos and
G.~Dvali, \PLB{436}{98}{257}.
%
\bibitem{pioline} I.~Antoniadis and B.~Pioline, 
\texttt{hep-ph/9902055};
K.~Benakli and Y.~Oz, \texttt{hep-th/9910090}.
%
\bibitem{bennett} S.C.~Bennett and C.E.~Wieman, \PRL{82}{99}{2484}.
\bibitem{qw} R.~Casalbuoni, S.~De Curtis, D.~Dominici and R.~Gatto,
\PLB{460}{99}{135}; J.L.~Rosner, \texttt{hep-ph/9907524};
J.~Erler and P.~Langacker, \texttt{hep-ph/9910315}.
%
\bibitem{ny}
P.~Nath and M.~Yamaguchi, \texttt{hep-ph/9902323}.
%
\bibitem{mp}
M.~Masip and A.~Pomarol, \PRD{60}{99}{096005}.
%
\bibitem{rw}
T.G.~Rizzo and J.D.~Wells,
\texttt{hep-ph/9906234}.
\bibitem{scc}
W.J.~Marciano, \PRD{60}{99}{093006};
A.~Strumia,
\texttt{hep-ph/9906266};
R.~Casalbuoni, S.~De Curtis, D.~Dominici and R.~Gatto,
\texttt{hep-ph/9907355};
C.D.~Carone,
\texttt{hep-ph/9907362};
F.~Cornet, M.~Relano and J.~Rico,
\texttt{hep-ph/9908299}.
%
\bibitem{alex} A.~Pomarol and M.~Quir\'os, \PLB{438}{98}{255};
I.~Antoniadis, S.~Dimopoulos, A.~Pomarol and M.~Quir\'os, \NPB{544}{99}{503};
A.~Delgado, A.~Pomarol and M.~Quir\'os, \PRD{60}{99}{095008}.
%
\bibitem{peskin} 
E.A.~Mirabelli and M.~Peskin, \PRD{58}{98}{065002}.  
%
\bibitem{PDG} C.~Caso et al., \EPJC{3}{98}{1};
J.~Mnich, talk presented at EPS-HEP 99, Tampere (Finland),
http://neutrino.pc.helsinki.fi/hep99/transparencies/Plenary/.
%
\bibitem{MS} W.J.~Marciano and A.~Sirlin, \PRD{35}{87}{1672}.
%
\bibitem{Feldman} G.J.~Feldman and R.D.~Cousins, \PRD{57}{98}{3873}.
%
\bibitem{as}
N.~Arkani-Hamed and M.~Schmaltz,
\texttt{hep-ph/9903417}.
%
\bibitem{st}
G.~Shiu and S.H.~Tye,
\PRD{58}{99}{106007}.
%
\bibitem{KKc}
M.~Bando, T.~Kugo, T.~Noguchi and K.~Yoshioka,
\texttt{hep-ph/9906549}; J.~Hisano and N.~Okada, \texttt{hep-ph/9909555}.
%
\bibitem{buras}
G.~Buchalla, A.J.~Buras and M.E.~Lautenbacher,
\RMP{68}{96}{1125}.
%
\bibitem{KKpro}
I.~Antoniadis, K.~Benakli and M.~Quiros,
\texttt{hep-ph/9905311};
P.~Nath, Y.~Yamada and M.~Yamaguchi,
\texttt{hep-ph/9905415};
T.G.~Rizzo,
\texttt{hep-ph/9909232}.
%
\end{thebibliography}
\end{document}